\documentclass[prd,a4paper,twocolumn,preprintnumbers,nofootinbib,superscriptaddress]{revtex4}

\pdfoutput=1
\usepackage[english]{babel}
\usepackage{amsmath,amssymb,amsfonts, bm,bbm,slashed, subdepth}
\usepackage{graphicx}
\usepackage[breaklinks=true,colorlinks=true,linkcolor=blue,urlcolor=blue,citecolor=blue]{hyperref}
\usepackage{enumerate}
\usepackage{setspace}
\usepackage{booktabs, tabularx}
\usepackage{units}
\usepackage{color}
\usepackage{multirow}
\usepackage[dvipsnames]{xcolor}
\usepackage[pscoord]{eso-pic}
\usepackage[normalem]{ulem}

\makeindex

\begin{document}
\preprint{ULB-TH/20-10}

\title{The role of CMB spectral distortions in the Hubble tension: a proof of principle}

\author{Matteo Lucca}
\email{mlucca@ulb.ac.be}
\affiliation{Service de Physique Th\'eorique, Universit\'e Libre de Bruxelles, Boulevard du Triomphe, CP225, 1050 Brussels, Belgium}

\begin{abstract}
	Although both early and late-time modifications of the $\Lambda$CDM model have been proposed to address the Hubble tension, compelling arguments suggest that for a solution to be successful it needs to modify the expansion history of the universe prior to recombination. This greatly increases the importance of precise CMB observations, and in this letter we make the argument for CMB spectral distortions, highlighting their potential role in constraining models that introduce significant shifts in the standard $\Lambda$CDM parameters, such as the scalar spectral index, in attempt to solve the Hubble tension.
\end{abstract}

\maketitle

In the last decade we have witnessed the emergence of an increasingly strong tension between the value of the local expansion rate of the universe, $H_0$, inferred from the early-time measurement of the Cosmic Microwave Background (CMB) radiation by Planck, ${H_0=67.4\pm0.5}$~km/(s~Mpc) \cite{Aghanim2018PlanckVI}, and the one reported by the SH0ES collaboration based on the observation of the local environment by the Hubble Space Telescope~(HST), ${H_0=74.03\pm1.42}$~km/(s~Mpc)~\cite{Riess2019Large} (see e.g., Fig. 17 of \cite{Freedman:2019jwv} for a nice historical overview on the evolution of this tension). Furthermore, in recent years also a variety of other cosmological probes, ranging from the combination of Baryon Acoustic Oscillation (BAO) and Big Bang Nucleosynthesis observations \cite{Cuceu:2019for} to lensed quasars \cite{Wong:2019kwg}, have helped consistently provide evidence for the presence of this tension (see e.g., Fig. 1 of \cite{Verde:2019ivm} and related text for a complete overview).

Although other minor tensions between the currently available observations exist (see e.g., \cite{Verde:2019ivm} again), this discrepancy is particularly interesting in the cosmological context as it represents the first considerable inconsistency (above 4$\sigma$) of the otherwise extremely successful $\Lambda$CDM model. As such, it motivated the development of a number of extensions of the $\Lambda$CDM model that could, in principle, alleviate this $H_0$ tension. In this regard, one can define two very broad classes of models: those that attempt to modify the history of the universe at early times, i.e., prior to recombination, and those that increase the Hubble rate at late times. Examples of the former class are Early Dark Energy (EDE) (see e.g., \cite{Poulin2018Early, Smith:2019ihp, Hill:2020osr, Ivanov:2020ril, Klypin:2020tud} as well as \cite{Lin:2019qug}), self-interacting neutrinos (SI$\nu$) (see e.g., \cite{Lancaster2017Tale, Kreisch2019Neutrino, Blinov:2019gcj, Escudero2019CMB, Escudero:2020hkf}) as well as models including additional light relics (see e.g.,~\cite{Agrawal2019Rock}), decaying dark matter (DM) (see e.g., \cite{Pandey:2019plg}), scenarios allowing for extra radiation (see e.g., \cite{Bernal2016Trouble, Schoeneberg2019BAO, Berghaus:2019cls}), and models that involve DM-dark radiation interactions (see e.g., \cite{BuenAbad2017Interacting, Archidiacono2019Constraining}). Very recently, also primordial magnetic fields (PMFs) have been shown to successfully address the tension~\cite{Jedamzik:2020krr}. On the other hand, popular late-time models involve other types of decaying DM (see e.g., \cite{Vattis:2019efj}), interactions between DM and Dark Energy~(DE) (see e.g., \cite{DiValentino2017Interacting, DiValentino2019Minimal, Lucca:2020zjb}), modified gravity (see e.g.,~\cite{Ade:2015rim} and references therein as well as \cite{Desmond:2019ygn}) and alternative parametrizations of the DE fluid (see e.g., \cite{Raveri:2019mxg, Keeley:2019esp, Alestas:2020mvb}). Of course, several other scenarios have been proposed that do not fall under this classification, such as e.g., \cite{DiValentino:2016hlg, Joudaki:2017zhq, Hart:2019dxi}.

However, it has been convincingly argued in a series of works \cite{Bernal2016Trouble, Poulin2018Implications, Aylor:2018drw, Knox2019Hubble} that models attempting to modify the expansion history of the universe at late times are incompatible with the combination of BAO and Supernovae Ia data (see e.g., Figs. 2 and 3 of \cite{Poulin2018Implications} for a graphical depiction). The reason for this is that the $H_0$ tension can be transposed in a physically more informative tension in the $H_0-r_s$ plane, where $r_s$ is the sound horizon at recombination time. Then, as can be seen e.g., in Fig.~13 of \cite{Bernal2016Trouble}, in this new plane the deviation from the $\Lambda$CDM prediction needs to happen in both directions, i.e., both $H_0$ and $r_s$ have to be modified by a given model in order to successfully solve the tension. However, as $r_s$ is only affected by the history of the universe prior to recombination, late-time modifications of the $\Lambda$CDM model are intrinsically incapable of fully solving the $H_0-r_s$ tension.

This idea points towards the fact that scenarios predicting early-time deviations from $\Lambda$CDM seem to be favored as possible solutions of the $H_0-r_s$ tension. Therefore, in order to better evaluate the additional free parameters introduced by the given models and distinguish between them, future CMB probes with improved sensitivities will prove crucial. For instance, as already argued in \cite{Smith:2019ihp}, the advent of CMB-S4 \cite{Abazajian2016CMB,Abazajian2019CMB} will allow to unambiguously confirm or rule out the presence of several EDE models. Also, as evident from Fig. 6 of \cite{Kreisch2019Neutrino}, the inclusion of currently available polarization data already enables to tighten the bounds on the region of the characteristic bi-modal posterior distribution that is able to alleviate the $H_0-r_s$ tension in the particular SI$\nu$ model considered there. Therefore, up-coming more advanced CMB experiments, such as the Simons Observatory \cite{Ade2018Simons}, will significantly help to test these types of scenarios \cite{Park:2019ibn}.

With this in mind, in this letter we discuss the role that CMB spectral distortions (SDs) will be able to play in constraining models that attempt to solve the Hubble tension by affecting the expansion history of the universe prior to recombination and thereby introduce significant shifts in the best-fit values of the $\Lambda$CDM parameters, such as the scalar spectral index. For further discussions on possible applications of SDs to the $H_0$ tension see~\cite{Abitbol:2019ewx}.

CMB SDs (see e.g., \cite{Zeldovich1969Interaction,Sunyaev1970Interaction, Burigana1991Formation, Hu1993ThermalizationI, Hu1993thermalizationII, Hu1995Wandering, Chluba2011Evolution, Chluba2013Distinguishing, Chluba2014Teasing, Chluba2016Which, Lucca2019Synergy, Chluba2019Spectral, Chluba2019Voyage, Fu:2020wkq}) are deviations of the CMB energy spectrum from a pure black body shape. They are formed prior to recombination\footnote{Note that SDs are created also at late times by the Sunyaev-Zeldovich effect \cite{Chluba2012Fast}, but we will not discuss these contributions any further within this work.} (up to redshifts $z\simeq2\times 10^6$) whenever energy is injected or extracted from the photon bath, or when the number density of the CMB photons is modified. Therefore, they are predicted to exist even within the $\Lambda$CDM model due to the presence of effects such as the adiabatic cooling of electrons and baryons, and the dissipation of acoustic waves. As shown for instance in Fig.~1 of~\cite{Fu:2020wkq}, the $\Lambda$CDM parameter to which SDs are most sensitive is the scalar spectral index $n_s$, followed by the amplitude of the primordial power spectrum (PPS) $A_s$, with only a mild dependence on the baryon and DM energy densities $\omega_b$ and $\omega_{\rm cdm}$. Via this strong dependence on the characteristics of the PPS, SDs are a particularly suited probe to test a variety of inflationary scenarios over scales completely different and complementary to those already constrained by CMB anisotropy measurements \cite{Sunyaev1970SmallII, Daly1991Spectral, Barrow1991Primordial, Hu1994Power, Chluba2012CMB, Chluba2012Inflaton, Cabass2016Distortions, Cabass2016Constraints, Fu:2020wkq}. Furthermore, SDs can be very sensitive also to energy injections from exotic models such as decaying DM \cite{Hu1993thermalizationII, Lucca2019Synergy, Acharya2019CMB} and evaporating primordial black holes \cite{Tashiro2008Constraints, Lucca2019Synergy, Acharya:2020jbv}, as well as to interactions between DM and baryons or photons \cite{AliHaimoud2015Constraints, Slatyer2018Early}, Axion-Like Particles \cite{Mukherjee:2018oeb, Mukherjee:2019dsu}, and PMFs \cite{Jedamzik:1999bm, Kunze:2013uja, Jedamzik:2018itu}. 

There is, however, another class of models that could introduce deviations in the observed SD signal from the $\Lambda$CDM prediction, and it will be the focus of this work. In fact, considering for instance the case of EDE, although this model does not directly create any non-standard contribution to the $\Lambda$CDM SD signal, significant shifts in several of the $\Lambda$CDM parameters are to be expected when the sound horizon is modified by the introduction of an additional DE-like component prior to recombination (see e.g., the introduction of \cite{Hill:2020osr} for a detailed discussion as well as Table I and Fig. 3 of \cite{Smith:2019ihp} for a quantitative and graphical overview). Indeed, the EDE fluid tends to suppress the growth of perturbations when it is non-negligible in the energy budget of the universe and, in order to preserve the accuracy of the fit to CMB anisotropy data, this needs to be compensated by a combination of an increased DM component and scalar spectral index. Eventually, this introduces strong degeneracies between e.g., $n_s$ and the EDE parameters such as the fraction of energy density of the EDE component $f_{\rm EDE}$. Therefore, since SDs are very sensitive to changes in $n_s$, they can also effectively constrain the specific EDE parameters by breaking the aforementioned degeneracies, and ultimately test the model's ability to solve the Hubble tension. 

This type of behavior is by no means exclusive of EDE, but applies, for instance, also to the SI$\nu$ model considered in \cite{Kreisch2019Neutrino}, as evident from Fig. 6 therein, as well as to the interacting majoron-neutrino scenario presented in \cite{Escudero2019CMB, Escudero:2020hkf} (see e.g., Table I of \cite{Escudero2019CMB}). Also models including evolving scalar fields such as the one discussed in \cite{Agrawal2019Rock} display similar features, as clear from Tab. 3 and Fig. 6 therein.

Thus, in order to concretely quantify the constraining power of SDs for this class of models, here we consider as a test-case the EDE scenario. As already mentioned above, the effects that this model has on the $\Lambda$CDM parameters are typically shared by many other scenarios, making the discussion presented here a proof of principle meant as general statement on the constraining potential of SDs, rather than the specific application to a single scenario. For this reason, we also neglect possible model-dependent effects that could cause non-standard deviations in the SD signal and only focus on the $\Lambda$CDM contributions. Moreover, this choice also goes beyond the ongoing debate on whether Large-Scale Structure data already disfavors the EDE scenario or not, see e.g., \cite{Hill:2020osr, Ivanov:2020ril} and \cite{Poulin2018Early, Smith:2019ihp, Klypin:2020tud} respectively, since here we focus purely on the features that the model imprints on the cosmological parameters, and not on the model itself.

For the numerical evaluation of the different cosmological quantities affected by EDE, such as the expansion rate and the perturbation equations, we incorporate the implementation of the EDE model presented in \cite{Hill:2020osr} in the version 3.0 of the Boltzmann solver \textsc{class} \cite{Lesgourgues2011Cosmic, Blas2011Cosmic}, which now allows for the computation of SDs as recently presented in \cite{Lucca2019Synergy, Fu:2020wkq}. For the SDs signal (henceforth also referred to as $\Lambda$CDM prediction) we include only the dissipation of acoustic waves and the adiabatic cooling of baryons and~electrons.
\begin{figure}
	\centering
	\includegraphics[width=\columnwidth]{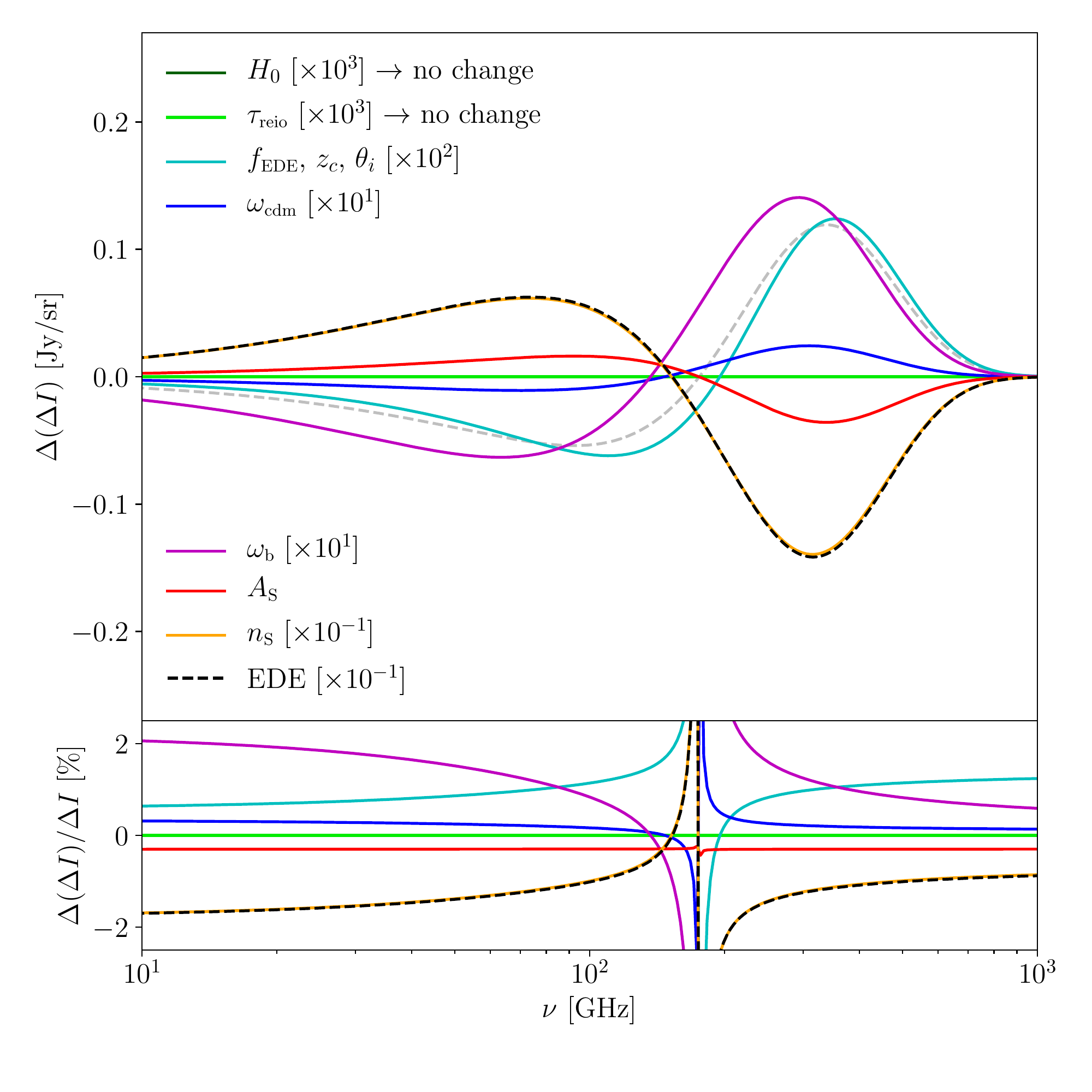}
	\caption{Deviations from the SD signal predicted by $\Lambda$CDM caused by assuming EDE as a cosmological model. The top panel shows the difference $\Delta(\Delta I)=\Delta I_{\rm \Lambda CDM}-\Delta I_{\rm  EDE}$ between the $\Lambda$CDM distortion spectrum $\Delta I_{\rm \Lambda CDM}$ computed assuming the mean values given in the left column of Tab. I of \cite{Smith:2019ihp} and the SD spectrum $\Delta I_{\rm  EDE}$ obtained by fixing a given cosmological parameter (or a combination of them) to the corresponding mean value obtained for the EDE model as given in the middle column of Tab. I of \cite{Smith:2019ihp} ($n=3$) while keeping all the others fixed to the $\Lambda$CDM values. The case labeled ``EDE" (black dashed line) assumes all parameters to follow the EDE mean values. The curves in both panels are multiplied by the factor reported in brackets in the legend. The $\Lambda$CDM spectrum, $\Delta I_{\rm \Lambda CDM}$, is shown in dashed gray multiplied by a factor of $10^{-2}$ as a reference. The bottom panel shows the same curves as the top panel, but in relative units.}
	\label{fig:EDE_test}
\end{figure}

With this combination of \textsc{class} versions it is first of all very instructive to compute the deviations from the SD signal predicted by $\Lambda$CDM caused by assuming EDE as a cosmological model. We show the results in Fig.~\ref{fig:EDE_test}. There, we assume for the $\Lambda$CDM model the parameter set given by the mean values of the left column of Tab. I of~\cite{Smith:2019ihp}, and change in turn every given parameter (leaving the others fixed to the $\Lambda$CDM values) according to the EDE mean values given in the middle column of Tab.~I of~\cite{Smith:2019ihp} ($n=3$). This allows to follow the impact that the variation of each parameter individually has on the SD signal. As it turns out, the strongest modification is given by the shift in the scalar spectral index $n_s$, which nearly perfectly overlaps with the ``EDE" prediction (black dashed line) computed assuming the full set of EDE mean values. Every other modification has only an impact of less than 0.5\%, either because the parameter itself is not affected by the model or because SDs are not sensitive to it. For instance, the increasing of the Hubble constant $H_0$, which is a key feature of the model, or the introduction of model-specific parameters such as $f_{\rm EDE}$ do not significantly affect the SD shape, as SDs are only sensitive to the early-time evolution of the universe (mostly before the onset of EDE, i.e., when $z>z_c$, where $z_c$ is the onset redshift of the EDE oscillations).

Note, however, that the fact that SDs are only mainly sensitive to one single cosmological parameter is not a problem. In fact, as already mentioned before, the inclusion of an SD probe would, in principle, significantly contribute to break the degeneracies involving the scalar spectral index, and thus indirectly also reduce the error bars on other parameters, such as $f_{\rm EDE}$, and in turn also~$H_0$.\footnote{This logic is very similar to how SDs could improve the bounds on the reionization optical depth $\tau_{\rm reio}$, although not directly sensitive to it, by breaking the well-known degeneracy this parameter shares with the amplitude of the PPS (see e.g., \cite{Fu:2020wkq} for additional comments).}

\begin{figure}
	\centering
	\includegraphics[width=\columnwidth]{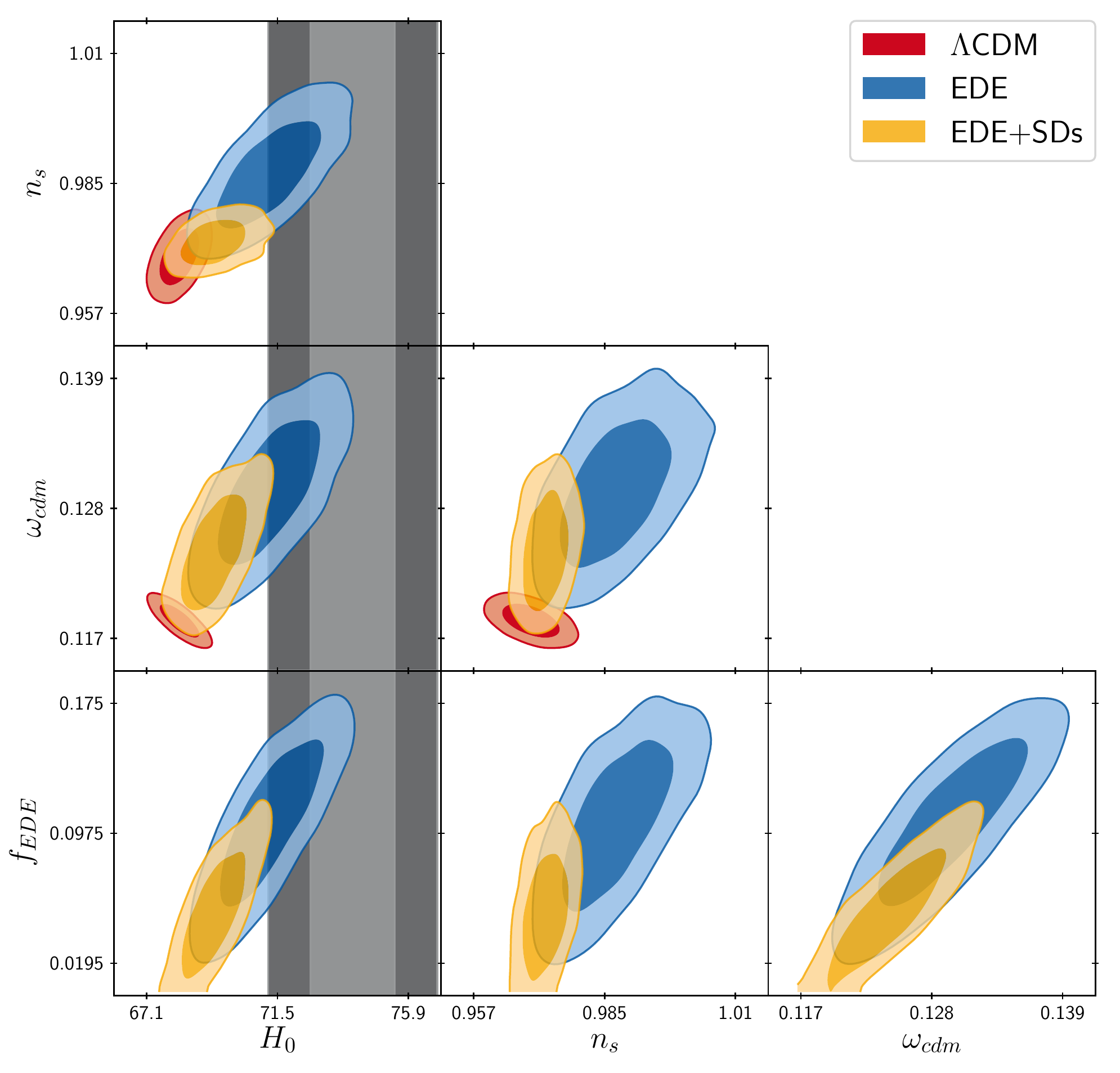}
	\caption{Posterior distributions (68\% and 95\% CL) of a selection of the cosmological parameters of the models considered in this work, i.e., $\Lambda$CDM (red) and EDE (blue and orange), for different combination of data sets, i.e., the same data sets listed in Sec. III A of \cite{Smith:2019ihp} (red and blue) and with the addition of the proposed SD mission PRISM (orange). The gray vertical bands correspond to the $H_0$ value reported by the SH0ES collaboration (68\% and 95\% CL).}
	\label{fig:MCMC_results}
\end{figure}

As these types of correlations can be quantitatively captured, for instance, by means of Monte Carlo Markov Chains (MCMCs) parameter scans sampling the different cosmological parameters involved, here we make use of the parameter extraction code \textsc{MontePython} \cite{Audren2013Conservative, Brinckmann2018MontePython} in combination with the \textsc{class} implementation already introduced before. In particular, we repeat the analysis already performed in~\cite{Smith:2019ihp} as baseline for the $\Lambda$CDM and EDE predictions (assuming $n=3$ for the EDE case, see Tab.~I and Fig.~3 therein). We adopt the very same conventions and use the same combination of Planck \cite{Aghanim2018PlanckVI}+ BAO \cite{Beutler2011Galaxy, Ross2014Clustering, Alam2016Clustering}+SH0ES \cite{Riess2019Large}+Supernovae \cite{Scolnic2017Complete} data as described in Sec. III A of the reference, with the only difference being that we employ Planck 2018 rather than Planck 2015 data. The results of our MCMC runs are listed in Tab.~\ref{tab:results} (first two columns) and the posterior distributions of a selection of the parameters most significant for our discussion are shown in Fig. \ref{fig:MCMC_results} (red and blue contours). The agreement with the results presented in Tab. I of \cite{Smith:2019ihp} is very good, as expected.

On the other hand, in order to forecast the impact of SDs we include a mock likelihood of the proposed SD mission PRISM \cite{Andre2014Prism} to the full sample of probes (see e.g., \cite{Perotto2006Probing, Brinckmann2018Promising, Lucca2019Synergy} for more details on mock likelihoods in general and on the numerical implementation). For this mission we assume a constant frequency resolution of 15~GHz in the range ${[30~\text{GHz} - 1~\text{THz}]}$, with an uncorrelated noise of ${5\times 10^{-27}~\text{W/m}^2\text{/Hz/sr}}$ for each resulting bin (one order of magnitude better than in the case of the PIXIE mission \cite{Kogut2011Primordial}). As a choice, we match the fiducial values to the mean values of the $\Lambda$CDM case presented in Tab. \ref{tab:results}, which corresponds to assuming that the hypothetical SD measurement perfectly confirms the $\Lambda$CDM prediction. In this way, the effect of adding an SD mission will result in both a reduction of the shifts in the means of the cosmological parameters when going from $\Lambda$CDM to EDE and a shrinking of the EDE posterior distributions.\footnote{Alternatively, one could have chosen as fiducial values the EDE means, which would have resulted only in a reduction of the uncertainties on the parameters, with no change in the mean values. The choice we make here has the advantage of highlighting more explicitly the role of SDs. We check, however, that the resulting uncertainties on the parameters involved are only marginally affected by the assumption made for the fiducial values.} 

As also done in \cite{Lucca2019Synergy, Fu:2020wkq}, in the SD likelihood we neglect galactic and extra-galactic foreground contaminations, which typically worsen the results by roughly one order of magnitude \cite{Abitbol2017Prospects}. In our case this is not a problem since it simply means that a more futuristic mission than PRISM, such as the recently proposed Voyage 2050 mission~\cite{Chluba2019Voyage}, would be able to achieve the same results. However, we do marginalize over the contributions from reionization and late-time sources, which is to say that we neglect the information stored in the ``primordial" $y$~distortions.

The results for the EDE run including the addition of the SD mission are listed in Tab. \ref{tab:results} (third column) and a selection of the most relevant parameters is shown in Fig.~\ref{fig:MCMC_results} (orange contours). By comparing the EDE posterior distributions with and without SDs it is immediately evident that, as expected, the degeneracy between $n_s$ and $f_{\rm EDE}$ is severely constrained. In turn, also the broadness of the $H_0$ contour is reduced to the point where it is not compatible with the SH0ES value any more (gray vertical bands in Fig.~\ref{fig:MCMC_results}). This suggests several interesting conclusions.
\begin{table}
	\centering
	\small
	\begin{tabular}{|c|c|c|c|}
		\hline\rule{0pt}{2.5ex} 
		Parameter 				& $\Lambda$CDM 					& EDE 								& EDE$+$SDs \\
		\hline\rule{0pt}{3.0ex}
		$100\, \omega_{b}$ 		& $2.253 \pm 0.014$ 			& $2.285_{-0.024}^{+0.022}$ 	 & $2.261 \pm 0.018$ \\
		$\omega_{\text {cdm }}$ & $0.1183_{-0.00091}^{+0.00089}$ & $0.1295_{-0.0040}^{+0.0039}$ 	 & $0.1245_{-0.0036}^{+0.0028}$ \\
		$100\, \theta_{s}$ 		& $1.0421 \pm 0.00029$ 			& $1.0415_{-0.00038}^{+0.00041}$ & $1.0417_{-0.00034}^{+0.00036}$ \\
		$10^{9}\, A_{s}$		& $2.121_{-0.034}^{+0.030}$ 	& $2.153_{-0.033}^{+0.034}$ 	 & $2.116_{-0.029}^{+0.028}$ \\
		$n_{s}$ 				& $0.9692 \pm 0.0038$ 			& $0.9873_{-0.0080}^{+0.0069}$ 	 & $0.9724_{-0.0032}^{+0.0030}$ \\
		$\tau_{\rm reio}$ 		& $0.0606_{-0.0081}^{+0.0071}$ 	& $0.0594_{-0.0085}^{+0.0073}$ 	 & $0.0538_{-0.0073}^{+0.0069}$ \\
		$\log_{10}(z_c)$	 	& - 							& $3.65_{-0.18}^{+0.13}$ 	 	 & $3.559_{-0.177}^{+0.078}$ \\
		$f_{\rm EDE}(z_c)$ 		& - 							& $0.103_{-0.029}^{+0.035}$ 	 & $< 0.095$ \\
		$\Theta_{i}$ 			& - 							& $2.640_{-0.010}^{+0.454}$ 	 & $> 0.40$ \\
		\hline \rule{0pt}{3.0ex}
		$H_{0}$ 				& $68.20 \pm 0.54$ 				& $71.2_{-1.1}^{+1.0}$ 		 & $69.41_{-0.75}^{+0.63}$ \\
		\hline
	\end{tabular}
	\caption{Mean and 1$\sigma$ uncertainty of the cosmological parameters of the models considered in this work, i.e., $\Lambda$CDM (first column) and EDE (second and third column), for different combination of data sets, i.e., the same data sets listed in Sec.~III~A of \cite{Smith:2019ihp} (first and second column) and with the addition of the proposed SD mission PRISM (third column). Upper and lower limits are given at 95\% CL.}
	\label{tab:results}
\end{table}

First of all, it fully confirms the aforementioned expectation that a future SD measurement will provide an additional powerful anchor to help constrain a whole class of models that introduce significant early-time modifications to the $\Lambda$CDM model (with particular emphasis to shifts in the scalar spectral index) in order to meaningfully address the $H_0-r_s$ tension and to avoid the very stringent CMB anisotropy constraints at the same time. Moreover, while this is certainly true with respect to current probes, as shown in Fig.~\ref{fig:MCMC_results}, the same conclusion also applies in relation to future CMB anisotropy observations. Indeed, although other upcoming missions such as \text{CMB-S4} \cite{Abazajian2016CMB,Abazajian2019CMB} would already be able to impose constraints on $H_0$ of similar magnitude to the ones presented in Tab. \ref{tab:results} for the EDE scenario (see Tab. VI of \cite{Smith:2019ihp} for a quantitative comparison), in this case SDs would still provide a completely independent, and thus very valuable, cosmological probe. This introduces an additional layer of importance to the synergy between CMB SDs and anisotropies \cite{Andre2014Prism, Chluba2014Teasing, Chluba2019Voyage, Lucca2019Synergy, Fu:2020wkq}. 

Secondly, the analysis presented here implies that an SD mission with PRISM-like sensitivities would already be enough to significantly contribute to the constraining of the aforementioned models, assuming a perfect control over foreground contaminations. Although it is very optimistic to assume a foreground removal to this degree of accuracy and fully accounting for late-time contaminations would greatly diminish the constraining power of PRISM (with an eventual close-to-complete overlap between orange and blue contours in Fig. \ref{fig:MCMC_results}), this obstacle could be overcome by means of more futuristic SD probes. For instance, as clear e.g., from Fig. 9 of~\cite{Chluba2019Voyage}, a mission such as Voyage 2050 would improve on the PRISM sensitivity by up to two orders of magnitude, which would be enough to compensate for the expected impact of the foreground contamination \cite{Abitbol2017Prospects}.

Furthermore, note that the aforementioned results completely depend on the choice made for the fiducial values of the SD likelihood. In fact, as already hinted in footnote 3, alternative choices for these parameters are possible and greater attention should be dedicated to the parameter uncertainties, rather than to the best fitting values. For instance, although a more careful investigation of the role of the fiducials and of the many available data set combinations is left for future work, preliminary results show that explicitly fixing the fiducial parameters to the EDE values presented in the middle column of Tab.~\ref{tab:results} results in an even stronger evidence for the presence of EDE when including SDs compared to the case without them, reducing the quoted error bars on $f_{\rm EDE}$ by roughly 30\%. Also the uncertainties on $H_0$ are reduced similarly to the values already presented in the right column of Tab.~\ref{tab:results}, preserving the compatibility with the SH0ES value. Therefore, in light of these considerations, the results presented in this letter should only be seen as one possible outcome of a future SD observation, bearing in mind that also alternative possibilities would prove to be rich of significant interpretations.

As a remark, it is also interesting to point out that future SD measurements will play an important role in another cosmological tension, the so-called $\sigma_8$ tension, which arises when comparing the matter power spectrum amplitude, $\sigma_8$, as predicted by CMB anisotropy measurements and Sunyaev-Zeldovich galaxy cluster analysis (see e.g., \cite{Douspis:2018xlj} and \cite{Bolliet:2018yaf} for recent overviews on the tension and the role of SDs, respectively). A more in-depth discussion about this aspect and dedicated sensitivity forecasts are however left for future work.

Finally, on a more general note, a key result of this letter is that SDs are an extremely versatile tool, able to probe cosmological scenarios well beyond straightforward energy injections in the CMB photon bath or modifications of the PPS. They can provide extremely valuable information on a vast variety of cosmological models and their future observation might uncover yet unknown physics, with relevance ranging from the very early to the very late universe.

\section*{Acknowledgments}
The author sincerely thanks Silvia Galli and Deanna C. Hooper for the many very insightful discussions, and Jens Chluba, Julien Lesgourgues, Vivian Poulin and Nils Sch\"oneberg for the helpful exchanges. The author also thanks Julien Lesgourgues and Nils Sch\"oneberg for kindly accepting to let the author use \textsc{class} v3.0 before the official release. This work is supported by the ``Probing dark matter with neutrinos" ULB-ARC convention and by the IISN convention 4.4503.15.

\newpage
\bibliographystyle{apsrev}
\bibliography{bibliography}

\begin{thebibliography}{93}
\expandafter\ifx\csname natexlab\endcsname\relax\def\natexlab#1{#1}\fi
\expandafter\ifx\csname bibnamefont\endcsname\relax
  \def\bibnamefont#1{#1}\fi
\expandafter\ifx\csname bibfnamefont\endcsname\relax
  \def\bibfnamefont#1{#1}\fi
\expandafter\ifx\csname citenamefont\endcsname\relax
  \def\citenamefont#1{#1}\fi
\expandafter\ifx\csname url\endcsname\relax
  \def\url#1{\texttt{#1}}\fi
\expandafter\ifx\csname urlprefix\endcsname\relax\def\urlprefix{URL }\fi
\providecommand{\bibinfo}[2]{#2}
\providecommand{\eprint}[2][]{\url{#2}}

\bibitem[{\citenamefont{Aghanim et~al.}(2018)}]{Aghanim2018PlanckVI}
\bibinfo{author}{\bibfnamefont{N.}~\bibnamefont{Aghanim}} \bibnamefont{et~al.}
  (\bibinfo{collaboration}{Planck}) (\bibinfo{year}{2018}),
  \eprint{1807.06209}.

\bibitem[{\citenamefont{Riess et~al.}(2019)\citenamefont{Riess, Casertano,
  Yuan, Macri, and Scolnic}}]{Riess2019Large}
\bibinfo{author}{\bibfnamefont{A.~G.} \bibnamefont{Riess}},
  \bibinfo{author}{\bibfnamefont{S.}~\bibnamefont{Casertano}},
  \bibinfo{author}{\bibfnamefont{W.}~\bibnamefont{Yuan}},
  \bibinfo{author}{\bibfnamefont{L.~M.} \bibnamefont{Macri}}, \bibnamefont{and}
  \bibinfo{author}{\bibfnamefont{D.}~\bibnamefont{Scolnic}},
  \textbf{\bibinfo{volume}{876}}, \bibinfo{pages}{85} (\bibinfo{year}{2019}),
  \eprint{1903.07603}.

\bibitem[{\citenamefont{Freedman et~al.}(2019)}]{Freedman:2019jwv}
\bibinfo{author}{\bibfnamefont{W.~L.} \bibnamefont{Freedman}}
  \bibnamefont{et~al.} (\bibinfo{year}{2019}), \eprint{1907.05922}.

\bibitem[{\citenamefont{Cuceu et~al.}(2019)\citenamefont{Cuceu, Farr, Lemos,
  and Font-Ribera}}]{Cuceu:2019for}
\bibinfo{author}{\bibfnamefont{A.}~\bibnamefont{Cuceu}},
  \bibinfo{author}{\bibfnamefont{J.}~\bibnamefont{Farr}},
  \bibinfo{author}{\bibfnamefont{P.}~\bibnamefont{Lemos}}, \bibnamefont{and}
  \bibinfo{author}{\bibfnamefont{A.}~\bibnamefont{Font-Ribera}},
  \bibinfo{journal}{JCAP} \textbf{\bibinfo{volume}{10}}, \bibinfo{pages}{044}
  (\bibinfo{year}{2019}), \eprint{1906.11628}.

\bibitem[{\citenamefont{Wong et~al.}(2019)}]{Wong:2019kwg}
\bibinfo{author}{\bibfnamefont{K.~C.} \bibnamefont{Wong}} \bibnamefont{et~al.}
  (\bibinfo{year}{2019}), \eprint{1907.04869}.

\bibitem[{\citenamefont{Verde et~al.}(2019)\citenamefont{Verde, Treu, and
  Riess}}]{Verde:2019ivm}
\bibinfo{author}{\bibfnamefont{L.}~\bibnamefont{Verde}},
  \bibinfo{author}{\bibfnamefont{T.}~\bibnamefont{Treu}}, \bibnamefont{and}
  \bibinfo{author}{\bibfnamefont{A.}~\bibnamefont{Riess}}
  (\bibinfo{year}{2019}), \eprint{1907.10625}.

\bibitem[{\citenamefont{Poulin et~al.}(2019)\citenamefont{Poulin, Smith,
  Karwal, and Kamionkowski}}]{Poulin2018Early}
\bibinfo{author}{\bibfnamefont{V.}~\bibnamefont{Poulin}},
  \bibinfo{author}{\bibfnamefont{T.~L.} \bibnamefont{Smith}},
  \bibinfo{author}{\bibfnamefont{T.}~\bibnamefont{Karwal}}, \bibnamefont{and}
  \bibinfo{author}{\bibfnamefont{M.}~\bibnamefont{Kamionkowski}},
  \bibinfo{journal}{Phys. Rev. Lett.} \textbf{\bibinfo{volume}{122}},
  \bibinfo{pages}{221301} (\bibinfo{year}{2019}), \eprint{1811.04083}.

\bibitem[{\citenamefont{Smith et~al.}(2020)\citenamefont{Smith, Poulin, and
  Amin}}]{Smith:2019ihp}
\bibinfo{author}{\bibfnamefont{T.~L.} \bibnamefont{Smith}},
  \bibinfo{author}{\bibfnamefont{V.}~\bibnamefont{Poulin}}, \bibnamefont{and}
  \bibinfo{author}{\bibfnamefont{M.~A.} \bibnamefont{Amin}},
  \bibinfo{journal}{Phys. Rev. D} \textbf{\bibinfo{volume}{101}},
  \bibinfo{pages}{063523} (\bibinfo{year}{2020}), \eprint{1908.06995}.

\bibitem[{\citenamefont{Hill et~al.}(2020)\citenamefont{Hill, McDonough,
  Toomey, and Alexander}}]{Hill:2020osr}
\bibinfo{author}{\bibfnamefont{J.~C.} \bibnamefont{Hill}},
  \bibinfo{author}{\bibfnamefont{E.}~\bibnamefont{McDonough}},
  \bibinfo{author}{\bibfnamefont{M.~W.} \bibnamefont{Toomey}},
  \bibnamefont{and} \bibinfo{author}{\bibfnamefont{S.}~\bibnamefont{Alexander}}
  (\bibinfo{year}{2020}), \eprint{2003.07355}.

\bibitem[{\citenamefont{Ivanov et~al.}(2020)}]{Ivanov:2020ril}
\bibinfo{author}{\bibfnamefont{M.~M.} \bibnamefont{Ivanov}}
  \bibnamefont{et~al.} (\bibinfo{year}{2020}), \eprint{2006.11235}.

\bibitem[{\citenamefont{Klypin et~al.}(2020)}]{Klypin:2020tud}
\bibinfo{author}{\bibfnamefont{A.}~\bibnamefont{Klypin}} \bibnamefont{et~al.}
  (\bibinfo{year}{2020}), \eprint{2006.14910}.

\bibitem[{\citenamefont{Lin et~al.}(2019)\citenamefont{Lin, Benevento, Hu, and
  Raveri}}]{Lin:2019qug}
\bibinfo{author}{\bibfnamefont{M.-X.} \bibnamefont{Lin}},
  \bibinfo{author}{\bibfnamefont{G.}~\bibnamefont{Benevento}},
  \bibinfo{author}{\bibfnamefont{W.}~\bibnamefont{Hu}}, \bibnamefont{and}
  \bibinfo{author}{\bibfnamefont{M.}~\bibnamefont{Raveri}},
  \bibinfo{journal}{Phys. Rev. D} \textbf{\bibinfo{volume}{100}},
  \bibinfo{pages}{063542} (\bibinfo{year}{2019}), \eprint{1905.12618}.

\bibitem[{\citenamefont{Lancaster et~al.}(2017)\citenamefont{Lancaster,
  Cyr-Racine, Knox, and Pan}}]{Lancaster2017Tale}
\bibinfo{author}{\bibfnamefont{L.}~\bibnamefont{Lancaster}},
  \bibinfo{author}{\bibfnamefont{F.-Y.} \bibnamefont{Cyr-Racine}},
  \bibinfo{author}{\bibfnamefont{L.}~\bibnamefont{Knox}}, \bibnamefont{and}
  \bibinfo{author}{\bibfnamefont{Z.}~\bibnamefont{Pan}},
  \bibinfo{journal}{JCAP} \textbf{\bibinfo{volume}{1707}}, \bibinfo{pages}{033}
  (\bibinfo{year}{2017}), \eprint{1704.06657}.

\bibitem[{\citenamefont{Kreisch et~al.}(2019)\citenamefont{Kreisch, Cyr-Racine,
  and Dor\'e}}]{Kreisch2019Neutrino}
\bibinfo{author}{\bibfnamefont{C.~D.} \bibnamefont{Kreisch}},
  \bibinfo{author}{\bibfnamefont{F.-Y.} \bibnamefont{Cyr-Racine}},
  \bibnamefont{and} \bibinfo{author}{\bibfnamefont{O.}~\bibnamefont{Dor\'e}}
  (\bibinfo{year}{2019}), \eprint{1902.00534}.

\bibitem[{\citenamefont{Blinov et~al.}(2019)\citenamefont{Blinov, Kelly,
  Krnjaic, and McDermott}}]{Blinov:2019gcj}
\bibinfo{author}{\bibfnamefont{N.}~\bibnamefont{Blinov}},
  \bibinfo{author}{\bibfnamefont{K.~J.} \bibnamefont{Kelly}},
  \bibinfo{author}{\bibfnamefont{G.~Z.} \bibnamefont{Krnjaic}},
  \bibnamefont{and} \bibinfo{author}{\bibfnamefont{S.~D.}
  \bibnamefont{McDermott}}, \bibinfo{journal}{Phys. Rev. Lett.}
  \textbf{\bibinfo{volume}{123}}, \bibinfo{pages}{191102}
  (\bibinfo{year}{2019}), \eprint{1905.02727}.

\bibitem[{\citenamefont{Escudero and Witte}(2019)}]{Escudero2019CMB}
\bibinfo{author}{\bibfnamefont{M.}~\bibnamefont{Escudero}} \bibnamefont{and}
  \bibinfo{author}{\bibfnamefont{S.~J.} \bibnamefont{Witte}}
  (\bibinfo{year}{2019}), \eprint{1909.04044}.

\bibitem[{\citenamefont{Escudero~Abenza and Witte}(2020)}]{Escudero:2020hkf}
\bibinfo{author}{\bibfnamefont{M.}~\bibnamefont{Escudero~Abenza}}
  \bibnamefont{and} \bibinfo{author}{\bibfnamefont{S.~J.} \bibnamefont{Witte}},
  in \emph{\bibinfo{booktitle}{{Prospects in Neutrino Physics}}}
  (\bibinfo{year}{2020}), \eprint{2004.01470}.

\bibitem[{\citenamefont{Agrawal et~al.}(2019)\citenamefont{Agrawal, Cyr-Racine,
  Pinner, and Randall}}]{Agrawal2019Rock}
\bibinfo{author}{\bibfnamefont{P.}~\bibnamefont{Agrawal}},
  \bibinfo{author}{\bibfnamefont{F.-Y.} \bibnamefont{Cyr-Racine}},
  \bibinfo{author}{\bibfnamefont{D.}~\bibnamefont{Pinner}}, \bibnamefont{and}
  \bibinfo{author}{\bibfnamefont{L.}~\bibnamefont{Randall}}
  (\bibinfo{year}{2019}), \eprint{1904.01016}.

\bibitem[{\citenamefont{Pandey et~al.}(2019)\citenamefont{Pandey, Karwal, and
  Das}}]{Pandey:2019plg}
\bibinfo{author}{\bibfnamefont{K.~L.} \bibnamefont{Pandey}},
  \bibinfo{author}{\bibfnamefont{T.}~\bibnamefont{Karwal}}, \bibnamefont{and}
  \bibinfo{author}{\bibfnamefont{S.}~\bibnamefont{Das}} (\bibinfo{year}{2019}),
  \eprint{1902.10636}.

\bibitem[{\citenamefont{Bernal et~al.}(2016)\citenamefont{Bernal, Verde, and
  Riess}}]{Bernal2016Trouble}
\bibinfo{author}{\bibfnamefont{J.~L.} \bibnamefont{Bernal}},
  \bibinfo{author}{\bibfnamefont{L.}~\bibnamefont{Verde}}, \bibnamefont{and}
  \bibinfo{author}{\bibfnamefont{A.~G.} \bibnamefont{Riess}},
  \bibinfo{journal}{JCAP} \textbf{\bibinfo{volume}{10}}, \bibinfo{pages}{019}
  (\bibinfo{year}{2016}), \eprint{1607.05617}.

\bibitem[{\citenamefont{Sch{\"o}neberg
  et~al.}(2019)\citenamefont{Sch{\"o}neberg, Lesgourgues, and
  Hooper}}]{Schoeneberg2019BAO}
\bibinfo{author}{\bibfnamefont{N.}~\bibnamefont{Sch{\"o}neberg}},
  \bibinfo{author}{\bibfnamefont{J.}~\bibnamefont{Lesgourgues}},
  \bibnamefont{and} \bibinfo{author}{\bibfnamefont{D.~C.}
  \bibnamefont{Hooper}}, \bibinfo{journal}{JCAP} \textbf{\bibinfo{volume}{10}},
  \bibinfo{pages}{029} (\bibinfo{year}{2019}), \eprint{1907.11594}.

\bibitem[{\citenamefont{Berghaus and Karwal}(2020)}]{Berghaus:2019cls}
\bibinfo{author}{\bibfnamefont{K.~V.} \bibnamefont{Berghaus}} \bibnamefont{and}
  \bibinfo{author}{\bibfnamefont{T.}~\bibnamefont{Karwal}},
  \bibinfo{journal}{Phys. Rev. D} \textbf{\bibinfo{volume}{101}},
  \bibinfo{pages}{083537} (\bibinfo{year}{2020}), \eprint{1911.06281}.

\bibitem[{\citenamefont{Buen-Abad et~al.}(2018)\citenamefont{Buen-Abad,
  Schmaltz, Lesgourgues, and Brinckmann}}]{BuenAbad2017Interacting}
\bibinfo{author}{\bibfnamefont{M.~A.} \bibnamefont{Buen-Abad}},
  \bibinfo{author}{\bibfnamefont{M.}~\bibnamefont{Schmaltz}},
  \bibinfo{author}{\bibfnamefont{J.}~\bibnamefont{Lesgourgues}},
  \bibnamefont{and}
  \bibinfo{author}{\bibfnamefont{T.}~\bibnamefont{Brinckmann}},
  \bibinfo{journal}{JCAP} \textbf{\bibinfo{volume}{1801}}, \bibinfo{pages}{008}
  (\bibinfo{year}{2018}), \eprint{1708.09406}.

\bibitem[{\citenamefont{Archidiacono et~al.}(2019)\citenamefont{Archidiacono,
  Hooper, Murgia, Bohr, Lesgourgues, and Viel}}]{Archidiacono2019Constraining}
\bibinfo{author}{\bibfnamefont{M.}~\bibnamefont{Archidiacono}},
  \bibinfo{author}{\bibfnamefont{D.~C.} \bibnamefont{Hooper}},
  \bibinfo{author}{\bibfnamefont{R.}~\bibnamefont{Murgia}},
  \bibinfo{author}{\bibfnamefont{S.}~\bibnamefont{Bohr}},
  \bibinfo{author}{\bibfnamefont{J.}~\bibnamefont{Lesgourgues}},
  \bibnamefont{and} \bibinfo{author}{\bibfnamefont{M.}~\bibnamefont{Viel}},
  \textbf{\bibinfo{volume}{1910}}, \bibinfo{pages}{055} (\bibinfo{year}{2019}),
  \eprint{1907.01496}.

\bibitem[{\citenamefont{Jedamzik and Pogosian}(2020)}]{Jedamzik:2020krr}
\bibinfo{author}{\bibfnamefont{K.}~\bibnamefont{Jedamzik}} \bibnamefont{and}
  \bibinfo{author}{\bibfnamefont{L.}~\bibnamefont{Pogosian}}
  (\bibinfo{year}{2020}), \eprint{2004.09487}.

\bibitem[{\citenamefont{Vattis et~al.}(2019)\citenamefont{Vattis, Koushiappas,
  and Loeb}}]{Vattis:2019efj}
\bibinfo{author}{\bibfnamefont{K.}~\bibnamefont{Vattis}},
  \bibinfo{author}{\bibfnamefont{S.~M.} \bibnamefont{Koushiappas}},
  \bibnamefont{and} \bibinfo{author}{\bibfnamefont{A.}~\bibnamefont{Loeb}},
  \bibinfo{journal}{Phys. Rev. D} \textbf{\bibinfo{volume}{99}},
  \bibinfo{pages}{121302} (\bibinfo{year}{2019}), \eprint{1903.06220}.

\bibitem[{\citenamefont{Di~Valentino et~al.}(2017)\citenamefont{Di~Valentino,
  Melchiorri, and Mena}}]{DiValentino2017Interacting}
\bibinfo{author}{\bibfnamefont{E.}~\bibnamefont{Di~Valentino}},
  \bibinfo{author}{\bibfnamefont{A.}~\bibnamefont{Melchiorri}},
  \bibnamefont{and} \bibinfo{author}{\bibfnamefont{O.}~\bibnamefont{Mena}},
  \bibinfo{journal}{Phys. Rev.} \textbf{\bibinfo{volume}{D96}},
  \bibinfo{pages}{043503} (\bibinfo{year}{2017}), \eprint{1704.08342}.

\bibitem[{\citenamefont{Di~Valentino et~al.}(2019)\citenamefont{Di~Valentino,
  Melchiorri, Mena, and Vagnozzi}}]{DiValentino2019Minimal}
\bibinfo{author}{\bibfnamefont{E.}~\bibnamefont{Di~Valentino}},
  \bibinfo{author}{\bibfnamefont{A.}~\bibnamefont{Melchiorri}},
  \bibinfo{author}{\bibfnamefont{O.}~\bibnamefont{Mena}}, \bibnamefont{and}
  \bibinfo{author}{\bibfnamefont{S.}~\bibnamefont{Vagnozzi}}
  (\bibinfo{year}{2019}), \eprint{1910.09853}.

\bibitem[{\citenamefont{Lucca and Hooper}(2020)}]{Lucca:2020zjb}
\bibinfo{author}{\bibfnamefont{M.}~\bibnamefont{Lucca}} \bibnamefont{and}
  \bibinfo{author}{\bibfnamefont{D.~C.} \bibnamefont{Hooper}}
  (\bibinfo{year}{2020}), \eprint{2002.06127}.

\bibitem[{\citenamefont{Ade et~al.}(2016)}]{Ade:2015rim}
\bibinfo{author}{\bibfnamefont{P.}~\bibnamefont{Ade}} \bibnamefont{et~al.}
  (\bibinfo{collaboration}{Planck}), \bibinfo{journal}{Astron. Astrophys.}
  \textbf{\bibinfo{volume}{594}}, \bibinfo{pages}{A14} (\bibinfo{year}{2016}),
  \eprint{1502.01590}.

\bibitem[{\citenamefont{Desmond et~al.}(2019)\citenamefont{Desmond, Jain, and
  Sakstein}}]{Desmond:2019ygn}
\bibinfo{author}{\bibfnamefont{H.}~\bibnamefont{Desmond}},
  \bibinfo{author}{\bibfnamefont{B.}~\bibnamefont{Jain}}, \bibnamefont{and}
  \bibinfo{author}{\bibfnamefont{J.}~\bibnamefont{Sakstein}},
  \bibinfo{journal}{Phys. Rev. D} \textbf{\bibinfo{volume}{100}},
  \bibinfo{pages}{043537} (\bibinfo{year}{2019}), \bibinfo{note}{[Erratum:
  Phys.Rev.D 101, 069904 (2020), Erratum: Phys.Rev.D 101, 129901 (2020)]},
  \eprint{1907.03778}.

\bibitem[{\citenamefont{Raveri}(2020)}]{Raveri:2019mxg}
\bibinfo{author}{\bibfnamefont{M.}~\bibnamefont{Raveri}},
  \bibinfo{journal}{Phys. Rev. D} \textbf{\bibinfo{volume}{101}},
  \bibinfo{pages}{083524} (\bibinfo{year}{2020}), \eprint{1902.01366}.

\bibitem[{\citenamefont{Keeley et~al.}(2019)\citenamefont{Keeley, Joudaki,
  Kaplinghat, and Kirkby}}]{Keeley:2019esp}
\bibinfo{author}{\bibfnamefont{R.~E.} \bibnamefont{Keeley}},
  \bibinfo{author}{\bibfnamefont{S.}~\bibnamefont{Joudaki}},
  \bibinfo{author}{\bibfnamefont{M.}~\bibnamefont{Kaplinghat}},
  \bibnamefont{and} \bibinfo{author}{\bibfnamefont{D.}~\bibnamefont{Kirkby}},
  \bibinfo{journal}{JCAP} \textbf{\bibinfo{volume}{12}}, \bibinfo{pages}{035}
  (\bibinfo{year}{2019}), \eprint{1905.10198}.

\bibitem[{\citenamefont{Alestas et~al.}(2020)\citenamefont{Alestas,
  Kazantzidis, and Perivolaropoulos}}]{Alestas:2020mvb}
\bibinfo{author}{\bibfnamefont{G.}~\bibnamefont{Alestas}},
  \bibinfo{author}{\bibfnamefont{L.}~\bibnamefont{Kazantzidis}},
  \bibnamefont{and}
  \bibinfo{author}{\bibfnamefont{L.}~\bibnamefont{Perivolaropoulos}},
  \bibinfo{journal}{Phys. Rev. D} \textbf{\bibinfo{volume}{101}},
  \bibinfo{pages}{123516} (\bibinfo{year}{2020}), \eprint{2004.08363}.

\bibitem[{\citenamefont{Di~Valentino et~al.}(2016)\citenamefont{Di~Valentino,
  Melchiorri, and Silk}}]{DiValentino:2016hlg}
\bibinfo{author}{\bibfnamefont{E.}~\bibnamefont{Di~Valentino}},
  \bibinfo{author}{\bibfnamefont{A.}~\bibnamefont{Melchiorri}},
  \bibnamefont{and} \bibinfo{author}{\bibfnamefont{J.}~\bibnamefont{Silk}},
  \bibinfo{journal}{Phys. Lett. B} \textbf{\bibinfo{volume}{761}},
  \bibinfo{pages}{242} (\bibinfo{year}{2016}), \eprint{1606.00634}.

\bibitem[{\citenamefont{Joudaki et~al.}(2018)\citenamefont{Joudaki, Kaplinghat,
  Keeley, and Kirkby}}]{Joudaki:2017zhq}
\bibinfo{author}{\bibfnamefont{S.}~\bibnamefont{Joudaki}},
  \bibinfo{author}{\bibfnamefont{M.}~\bibnamefont{Kaplinghat}},
  \bibinfo{author}{\bibfnamefont{R.}~\bibnamefont{Keeley}}, \bibnamefont{and}
  \bibinfo{author}{\bibfnamefont{D.}~\bibnamefont{Kirkby}},
  \bibinfo{journal}{Phys. Rev. D} \textbf{\bibinfo{volume}{97}},
  \bibinfo{pages}{123501} (\bibinfo{year}{2018}), \eprint{1710.04236}.

\bibitem[{\citenamefont{Hart and Chluba}(2020)}]{Hart:2019dxi}
\bibinfo{author}{\bibfnamefont{L.}~\bibnamefont{Hart}} \bibnamefont{and}
  \bibinfo{author}{\bibfnamefont{J.}~\bibnamefont{Chluba}},
  \bibinfo{journal}{Mon. Not. Roy. Astron. Soc.}
  \textbf{\bibinfo{volume}{493}}, \bibinfo{pages}{3255} (\bibinfo{year}{2020}),
  \eprint{1912.03986}.

\bibitem[{\citenamefont{Poulin et~al.}(2018)\citenamefont{Poulin, Boddy, Bird,
  and Kamionkowski}}]{Poulin2018Implications}
\bibinfo{author}{\bibfnamefont{V.}~\bibnamefont{Poulin}},
  \bibinfo{author}{\bibfnamefont{K.~K.} \bibnamefont{Boddy}},
  \bibinfo{author}{\bibfnamefont{S.}~\bibnamefont{Bird}}, \bibnamefont{and}
  \bibinfo{author}{\bibfnamefont{M.}~\bibnamefont{Kamionkowski}},
  \bibinfo{journal}{Phys. Rev. D} \textbf{\bibinfo{volume}{97}},
  \bibinfo{pages}{123504} (\bibinfo{year}{2018}), \eprint{1803.02474}.

\bibitem[{\citenamefont{Aylor et~al.}(2019)\citenamefont{Aylor, Joy, Knox,
  Millea, Raghunathan, and Wu}}]{Aylor:2018drw}
\bibinfo{author}{\bibfnamefont{K.}~\bibnamefont{Aylor}},
  \bibinfo{author}{\bibfnamefont{M.}~\bibnamefont{Joy}},
  \bibinfo{author}{\bibfnamefont{L.}~\bibnamefont{Knox}},
  \bibinfo{author}{\bibfnamefont{M.}~\bibnamefont{Millea}},
  \bibinfo{author}{\bibfnamefont{S.}~\bibnamefont{Raghunathan}},
  \bibnamefont{and} \bibinfo{author}{\bibfnamefont{W.~K.} \bibnamefont{Wu}},
  \bibinfo{journal}{Astrophys. J.} \textbf{\bibinfo{volume}{874}},
  \bibinfo{pages}{4} (\bibinfo{year}{2019}), \eprint{1811.00537}.

\bibitem[{\citenamefont{Knox and Millea}(2020)}]{Knox2019Hubble}
\bibinfo{author}{\bibfnamefont{L.}~\bibnamefont{Knox}} \bibnamefont{and}
  \bibinfo{author}{\bibfnamefont{M.}~\bibnamefont{Millea}},
  \bibinfo{journal}{Phys. Rev. D} \textbf{\bibinfo{volume}{101}},
  \bibinfo{pages}{043533} (\bibinfo{year}{2020}), \eprint{1908.03663}.

\bibitem[{\citenamefont{Abazajian et~al.}(2016)}]{Abazajian2016CMB}
\bibinfo{author}{\bibfnamefont{K.~N.} \bibnamefont{Abazajian}}
  \bibnamefont{et~al.} (\bibinfo{collaboration}{CMB-S4})
  (\bibinfo{year}{2016}), \eprint{1610.02743}.

\bibitem[{\citenamefont{Abazajian et~al.}(2019)}]{Abazajian2019CMB}
\bibinfo{author}{\bibfnamefont{K.}~\bibnamefont{Abazajian}}
  \bibnamefont{et~al.} (\bibinfo{year}{2019}), \eprint{1907.04473}.

\bibitem[{\citenamefont{Ade et~al.}(2019)}]{Ade2018Simons}
\bibinfo{author}{\bibfnamefont{P.}~\bibnamefont{Ade}} \bibnamefont{et~al.},
  \bibinfo{journal}{\jcap} \textbf{\bibinfo{volume}{2019}}, \bibinfo{eid}{056}
  (\bibinfo{year}{2019}), \eprint{1808.07445}.

\bibitem[{\citenamefont{Park et~al.}(2019)\citenamefont{Park, Kreisch, Dunkley,
  Hadzhiyska, and Cyr-Racine}}]{Park:2019ibn}
\bibinfo{author}{\bibfnamefont{M.}~\bibnamefont{Park}},
  \bibinfo{author}{\bibfnamefont{C.~D.} \bibnamefont{Kreisch}},
  \bibinfo{author}{\bibfnamefont{J.}~\bibnamefont{Dunkley}},
  \bibinfo{author}{\bibfnamefont{B.}~\bibnamefont{Hadzhiyska}},
  \bibnamefont{and} \bibinfo{author}{\bibfnamefont{F.-Y.}
  \bibnamefont{Cyr-Racine}}, \bibinfo{journal}{Phys. Rev. D}
  \textbf{\bibinfo{volume}{100}}, \bibinfo{pages}{063524}
  (\bibinfo{year}{2019}), \eprint{1904.02625}.

\bibitem[{\citenamefont{Abitbol et~al.}(2019)\citenamefont{Abitbol, Hill, and
  Chluba}}]{Abitbol:2019ewx}
\bibinfo{author}{\bibfnamefont{M.~H.} \bibnamefont{Abitbol}},
  \bibinfo{author}{\bibfnamefont{J.~C.} \bibnamefont{Hill}}, \bibnamefont{and}
  \bibinfo{author}{\bibfnamefont{J.}~\bibnamefont{Chluba}}
  (\bibinfo{year}{2019}), \eprint{1910.09881}.

\bibitem[{\citenamefont{Zeldovich and
  Sunyaev}(1969)}]{Zeldovich1969Interaction}
\bibinfo{author}{\bibfnamefont{Y.~B.} \bibnamefont{Zeldovich}}
  \bibnamefont{and} \bibinfo{author}{\bibfnamefont{R.}~\bibnamefont{Sunyaev}},
  \bibinfo{journal}{Astrophysics and Space Science}
  \textbf{\bibinfo{volume}{4}}, \bibinfo{pages}{301} (\bibinfo{year}{1969}).

\bibitem[{\citenamefont{Sunyaev and Zeldovich}(1970)}]{Sunyaev1970Interaction}
\bibinfo{author}{\bibfnamefont{R.~A.} \bibnamefont{Sunyaev}} \bibnamefont{and}
  \bibinfo{author}{\bibfnamefont{Y.~B.} \bibnamefont{Zeldovich}},
  \bibinfo{journal}{Astrophysics and Space Science}
  \textbf{\bibinfo{volume}{7}}, \bibinfo{pages}{20} (\bibinfo{year}{1970}),
  ISSN \bibinfo{issn}{1572-946X}.

\bibitem[{\citenamefont{Burigana et~al.}(1991)\citenamefont{Burigana, Danese,
  and De~Zotti}}]{Burigana1991Formation}
\bibinfo{author}{\bibfnamefont{C.}~\bibnamefont{Burigana}},
  \bibinfo{author}{\bibfnamefont{L.}~\bibnamefont{Danese}}, \bibnamefont{and}
  \bibinfo{author}{\bibfnamefont{G.}~\bibnamefont{De~Zotti}},
  \bibinfo{journal}{Astronomy and Astrophysics} \textbf{\bibinfo{volume}{246}},
  \bibinfo{pages}{49} (\bibinfo{year}{1991}).

\bibitem[{\citenamefont{Hu and
  Silk}(1993{\natexlab{a}})}]{Hu1993ThermalizationI}
\bibinfo{author}{\bibfnamefont{W.}~\bibnamefont{Hu}} \bibnamefont{and}
  \bibinfo{author}{\bibfnamefont{J.}~\bibnamefont{Silk}},
  \bibinfo{journal}{Phys. Rev.} \textbf{\bibinfo{volume}{D48}},
  \bibinfo{pages}{485} (\bibinfo{year}{1993}{\natexlab{a}}).

\bibitem[{\citenamefont{Hu and
  Silk}(1993{\natexlab{b}})}]{Hu1993thermalizationII}
\bibinfo{author}{\bibfnamefont{W.}~\bibnamefont{Hu}} \bibnamefont{and}
  \bibinfo{author}{\bibfnamefont{J.}~\bibnamefont{Silk}},
  \bibinfo{journal}{Phys. Rev. Lett.} \textbf{\bibinfo{volume}{70}},
  \bibinfo{pages}{2661} (\bibinfo{year}{1993}{\natexlab{b}}).

\bibitem[{\citenamefont{Hu}(1995)}]{Hu1995Wandering}
\bibinfo{author}{\bibfnamefont{W.~T.} \bibnamefont{Hu}}, Ph.D. thesis,
  \bibinfo{school}{UC, Berkeley} (\bibinfo{year}{1995}),
  \eprint{astro-ph/9508126}.

\bibitem[{\citenamefont{Chluba and Sunyaev}(2012)}]{Chluba2011Evolution}
\bibinfo{author}{\bibfnamefont{J.}~\bibnamefont{Chluba}} \bibnamefont{and}
  \bibinfo{author}{\bibfnamefont{R.~A.} \bibnamefont{Sunyaev}},
  \bibinfo{journal}{Mon. Not. Roy. Astron. Soc.}
  \textbf{\bibinfo{volume}{419}}, \bibinfo{pages}{1294} (\bibinfo{year}{2012}),
  \eprint{1109.6552}.

\bibitem[{\citenamefont{Chluba}(2013)}]{Chluba2013Distinguishing}
\bibinfo{author}{\bibfnamefont{J.}~\bibnamefont{Chluba}},
  \bibinfo{journal}{Mon. Not. Roy. Astron. Soc.}
  \textbf{\bibinfo{volume}{436}}, \bibinfo{pages}{2232} (\bibinfo{year}{2013}),
  \eprint{1304.6121}.

\bibitem[{\citenamefont{Chluba and Jeong}(2014)}]{Chluba2014Teasing}
\bibinfo{author}{\bibfnamefont{J.}~\bibnamefont{Chluba}} \bibnamefont{and}
  \bibinfo{author}{\bibfnamefont{D.}~\bibnamefont{Jeong}},
  \bibinfo{journal}{Mon. Not. Roy. Astron. Soc.}
  \textbf{\bibinfo{volume}{438}}, \bibinfo{pages}{2065} (\bibinfo{year}{2014}),
  \eprint{1306.5751}.

\bibitem[{\citenamefont{Chluba}(2016)}]{Chluba2016Which}
\bibinfo{author}{\bibfnamefont{J.}~\bibnamefont{Chluba}},
  \bibinfo{journal}{Mon. Not. Roy. Astron. Soc.}
  \textbf{\bibinfo{volume}{460}}, \bibinfo{pages}{227} (\bibinfo{year}{2016}),
  \eprint{1603.02496}.

\bibitem[{\citenamefont{Lucca et~al.}(2020)\citenamefont{Lucca, Sch{\"o}neberg,
  Hooper, Lesgourgues, and Chluba}}]{Lucca2019Synergy}
\bibinfo{author}{\bibfnamefont{M.}~\bibnamefont{Lucca}},
  \bibinfo{author}{\bibfnamefont{N.}~\bibnamefont{Sch{\"o}neberg}},
  \bibinfo{author}{\bibfnamefont{D.~C.} \bibnamefont{Hooper}},
  \bibinfo{author}{\bibfnamefont{J.}~\bibnamefont{Lesgourgues}},
  \bibnamefont{and} \bibinfo{author}{\bibfnamefont{J.}~\bibnamefont{Chluba}},
  \bibinfo{journal}{JCAP} \textbf{\bibinfo{volume}{02}}, \bibinfo{pages}{026}
  (\bibinfo{year}{2020}), \eprint{1910.04619}.

\bibitem[{\citenamefont{Chluba
  et~al.}(2019{\natexlab{a}})}]{Chluba2019Spectral}
\bibinfo{author}{\bibfnamefont{J.}~\bibnamefont{Chluba}} \bibnamefont{et~al.},
  \bibinfo{journal}{\baas} \textbf{\bibinfo{volume}{51}}, \bibinfo{eid}{184}
  (\bibinfo{year}{2019}{\natexlab{a}}), \eprint{1903.04218}.

\bibitem[{\citenamefont{Chluba et~al.}(2019{\natexlab{b}})}]{Chluba2019Voyage}
\bibinfo{author}{\bibfnamefont{J.}~\bibnamefont{Chluba}} \bibnamefont{et~al.}
  (\bibinfo{year}{2019}{\natexlab{b}}), \eprint{1909.01593}.

\bibitem[{\citenamefont{Fu et~al.}(2020)\citenamefont{Fu, Lucca, Galli,
  Battistelli, Hooper, Lesgourgues, and Schöneberg}}]{Fu:2020wkq}
\bibinfo{author}{\bibfnamefont{H.}~\bibnamefont{Fu}},
  \bibinfo{author}{\bibfnamefont{M.}~\bibnamefont{Lucca}},
  \bibinfo{author}{\bibfnamefont{S.}~\bibnamefont{Galli}},
  \bibinfo{author}{\bibfnamefont{E.~S.} \bibnamefont{Battistelli}},
  \bibinfo{author}{\bibfnamefont{D.~C.} \bibnamefont{Hooper}},
  \bibinfo{author}{\bibfnamefont{J.}~\bibnamefont{Lesgourgues}},
  \bibnamefont{and}
  \bibinfo{author}{\bibfnamefont{N.}~\bibnamefont{Schöneberg}}
  (\bibinfo{year}{2020}), \eprint{2006.12886}.

\bibitem[{\citenamefont{Chluba et~al.}(2012{\natexlab{a}})\citenamefont{Chluba,
  Nagai, Sazonov, and Nelson}}]{Chluba2012Fast}
\bibinfo{author}{\bibfnamefont{J.}~\bibnamefont{Chluba}},
  \bibinfo{author}{\bibfnamefont{D.}~\bibnamefont{Nagai}},
  \bibinfo{author}{\bibfnamefont{S.}~\bibnamefont{Sazonov}}, \bibnamefont{and}
  \bibinfo{author}{\bibfnamefont{K.}~\bibnamefont{Nelson}},
  \bibinfo{journal}{Mon. Not. Roy. Astron. Soc.}
  \textbf{\bibinfo{volume}{426}}, \bibinfo{pages}{510}
  (\bibinfo{year}{2012}{\natexlab{a}}), \eprint{1205.5778}.

\bibitem[{\citenamefont{{Sunyaev} and {Zeldovich}}(1970)}]{Sunyaev1970SmallII}
\bibinfo{author}{\bibfnamefont{R.~A.} \bibnamefont{{Sunyaev}}}
  \bibnamefont{and} \bibinfo{author}{\bibfnamefont{Y.~B.}
  \bibnamefont{{Zeldovich}}}, \bibinfo{journal}{\apss}
  \textbf{\bibinfo{volume}{9}}, \bibinfo{pages}{368} (\bibinfo{year}{1970}).

\bibitem[{\citenamefont{Daly}(1991)}]{Daly1991Spectral}
\bibinfo{author}{\bibfnamefont{R.}~\bibnamefont{Daly}}, \bibinfo{journal}{The
  Astrophysical Journal} \textbf{\bibinfo{volume}{371}}, \bibinfo{pages}{14}
  (\bibinfo{year}{1991}).

\bibitem[{\citenamefont{Barrow and Coles}(1991)}]{Barrow1991Primordial}
\bibinfo{author}{\bibfnamefont{J.~D.} \bibnamefont{Barrow}} \bibnamefont{and}
  \bibinfo{author}{\bibfnamefont{P.}~\bibnamefont{Coles}},
  \bibinfo{journal}{Monthly Notices of the Royal Astronomical Society}
  \textbf{\bibinfo{volume}{248}}, \bibinfo{pages}{52} (\bibinfo{year}{1991}),
  ISSN \bibinfo{issn}{0035-8711}.

\bibitem[{\citenamefont{{Hu} et~al.}(1994)\citenamefont{{Hu}, {Scott}, and
  {Silk}}}]{Hu1994Power}
\bibinfo{author}{\bibfnamefont{W.}~\bibnamefont{{Hu}}},
  \bibinfo{author}{\bibfnamefont{D.}~\bibnamefont{{Scott}}}, \bibnamefont{and}
  \bibinfo{author}{\bibfnamefont{J.}~\bibnamefont{{Silk}}},
  \bibinfo{journal}{\apjl} \textbf{\bibinfo{volume}{430}}, \bibinfo{pages}{L5}
  (\bibinfo{year}{1994}), \eprint{arXiv:astro-ph/9402045}.

\bibitem[{\citenamefont{Chluba et~al.}(2012{\natexlab{b}})\citenamefont{Chluba,
  Khatri, and Sunyaev}}]{Chluba2012CMB}
\bibinfo{author}{\bibfnamefont{J.}~\bibnamefont{Chluba}},
  \bibinfo{author}{\bibfnamefont{R.}~\bibnamefont{Khatri}}, \bibnamefont{and}
  \bibinfo{author}{\bibfnamefont{R.~A.} \bibnamefont{Sunyaev}},
  \bibinfo{journal}{Mon. Not. Roy. Astron. Soc.}
  \textbf{\bibinfo{volume}{425}}, \bibinfo{pages}{1129}
  (\bibinfo{year}{2012}{\natexlab{b}}), \eprint{1202.0057}.

\bibitem[{\citenamefont{{Chluba} et~al.}(2012)\citenamefont{{Chluba},
  {Erickcek}, and {Ben-Dayan}}}]{Chluba2012Inflaton}
\bibinfo{author}{\bibfnamefont{J.}~\bibnamefont{{Chluba}}},
  \bibinfo{author}{\bibfnamefont{A.~L.} \bibnamefont{{Erickcek}}},
  \bibnamefont{and}
  \bibinfo{author}{\bibfnamefont{I.}~\bibnamefont{{Ben-Dayan}}},
  \bibinfo{journal}{\apj} \textbf{\bibinfo{volume}{758}}, \bibinfo{eid}{76}
  (\bibinfo{year}{2012}), \eprint{1203.2681}.

\bibitem[{\citenamefont{{Cabass}
  et~al.}(2016{\natexlab{a}})\citenamefont{{Cabass}, {Melchiorri}, and
  {Pajer}}}]{Cabass2016Distortions}
\bibinfo{author}{\bibfnamefont{G.}~\bibnamefont{{Cabass}}},
  \bibinfo{author}{\bibfnamefont{A.}~\bibnamefont{{Melchiorri}}},
  \bibnamefont{and} \bibinfo{author}{\bibfnamefont{E.}~\bibnamefont{{Pajer}}},
  \bibinfo{journal}{Phys.~Rev.} \textbf{\bibinfo{volume}{D93}},
  \bibinfo{eid}{083515} (\bibinfo{year}{2016}{\natexlab{a}}),
  \eprint{1602.05578}.

\bibitem[{\citenamefont{{Cabass}
  et~al.}(2016{\natexlab{b}})\citenamefont{{Cabass}, {Di Valentino},
  {Melchiorri}, {Pajer}, and {Silk}}}]{Cabass2016Constraints}
\bibinfo{author}{\bibfnamefont{G.}~\bibnamefont{{Cabass}}},
  \bibinfo{author}{\bibfnamefont{E.}~\bibnamefont{{Di Valentino}}},
  \bibinfo{author}{\bibfnamefont{A.}~\bibnamefont{{Melchiorri}}},
  \bibinfo{author}{\bibfnamefont{E.}~\bibnamefont{{Pajer}}}, \bibnamefont{and}
  \bibinfo{author}{\bibfnamefont{J.}~\bibnamefont{{Silk}}},
  \bibinfo{journal}{Phys.~Rev.} \textbf{\bibinfo{volume}{D94}},
  \bibinfo{eid}{023523} (\bibinfo{year}{2016}{\natexlab{b}}).

\bibitem[{\citenamefont{Acharya and Khatri}(2019)}]{Acharya2019CMB}
\bibinfo{author}{\bibfnamefont{S.~K.} \bibnamefont{Acharya}} \bibnamefont{and}
  \bibinfo{author}{\bibfnamefont{R.}~\bibnamefont{Khatri}},
  \bibinfo{journal}{JCAP} \textbf{\bibinfo{volume}{12}}, \bibinfo{pages}{046}
  (\bibinfo{year}{2019}), \eprint{1910.06272}.

\bibitem[{\citenamefont{Tashiro and Sugiyama}(2008)}]{Tashiro2008Constraints}
\bibinfo{author}{\bibfnamefont{H.}~\bibnamefont{Tashiro}} \bibnamefont{and}
  \bibinfo{author}{\bibfnamefont{N.}~\bibnamefont{Sugiyama}},
  \bibinfo{journal}{Phys. Rev.} \textbf{\bibinfo{volume}{D78}},
  \bibinfo{pages}{023004} (\bibinfo{year}{2008}), \eprint{0801.3172}.

\bibitem[{\citenamefont{Acharya and Khatri}(2020)}]{Acharya:2020jbv}
\bibinfo{author}{\bibfnamefont{S.~K.} \bibnamefont{Acharya}} \bibnamefont{and}
  \bibinfo{author}{\bibfnamefont{R.}~\bibnamefont{Khatri}},
  \bibinfo{journal}{JCAP} \textbf{\bibinfo{volume}{06}}, \bibinfo{pages}{018}
  (\bibinfo{year}{2020}), \eprint{2002.00898}.

\bibitem[{\citenamefont{Ali-Ha\"{\i}moud
  et~al.}(2015)\citenamefont{Ali-Ha\"{\i}moud, Chluba, and
  Kamionkowski}}]{AliHaimoud2015Constraints}
\bibinfo{author}{\bibfnamefont{Y.}~\bibnamefont{Ali-Ha\"{\i}moud}},
  \bibinfo{author}{\bibfnamefont{J.}~\bibnamefont{Chluba}}, \bibnamefont{and}
  \bibinfo{author}{\bibfnamefont{M.}~\bibnamefont{Kamionkowski}},
  \bibinfo{journal}{Phys. Rev. Lett.} \textbf{\bibinfo{volume}{115}},
  \bibinfo{pages}{071304} (\bibinfo{year}{2015}).

\bibitem[{\citenamefont{Slatyer and Wu}(2018)}]{Slatyer2018Early}
\bibinfo{author}{\bibfnamefont{T.~R.} \bibnamefont{Slatyer}} \bibnamefont{and}
  \bibinfo{author}{\bibfnamefont{C.-L.} \bibnamefont{Wu}},
  \bibinfo{journal}{Phys. Rev. D} \textbf{\bibinfo{volume}{98}},
  \bibinfo{pages}{023013} (\bibinfo{year}{2018}).

\bibitem[{\citenamefont{Mukherjee et~al.}(2018)\citenamefont{Mukherjee, Khatri,
  and Wandelt}}]{Mukherjee:2018oeb}
\bibinfo{author}{\bibfnamefont{S.}~\bibnamefont{Mukherjee}},
  \bibinfo{author}{\bibfnamefont{R.}~\bibnamefont{Khatri}}, \bibnamefont{and}
  \bibinfo{author}{\bibfnamefont{B.~D.} \bibnamefont{Wandelt}},
  \bibinfo{journal}{JCAP} \textbf{\bibinfo{volume}{04}}, \bibinfo{pages}{045}
  (\bibinfo{year}{2018}), \eprint{1801.09701}.

\bibitem[{\citenamefont{Mukherjee et~al.}(2020)\citenamefont{Mukherjee,
  Spergel, Khatri, and Wandelt}}]{Mukherjee:2019dsu}
\bibinfo{author}{\bibfnamefont{S.}~\bibnamefont{Mukherjee}},
  \bibinfo{author}{\bibfnamefont{D.~N.} \bibnamefont{Spergel}},
  \bibinfo{author}{\bibfnamefont{R.}~\bibnamefont{Khatri}}, \bibnamefont{and}
  \bibinfo{author}{\bibfnamefont{B.~D.} \bibnamefont{Wandelt}},
  \bibinfo{journal}{JCAP} \textbf{\bibinfo{volume}{02}}, \bibinfo{pages}{032}
  (\bibinfo{year}{2020}), \eprint{1908.07534}.

\bibitem[{\citenamefont{Jedamzik et~al.}(2000)\citenamefont{Jedamzik,
  Katalinic, and Olinto}}]{Jedamzik:1999bm}
\bibinfo{author}{\bibfnamefont{K.}~\bibnamefont{Jedamzik}},
  \bibinfo{author}{\bibfnamefont{V.}~\bibnamefont{Katalinic}},
  \bibnamefont{and} \bibinfo{author}{\bibfnamefont{A.~V.}
  \bibnamefont{Olinto}}, \bibinfo{journal}{Phys. Rev. Lett.}
  \textbf{\bibinfo{volume}{85}}, \bibinfo{pages}{700} (\bibinfo{year}{2000}),
  \eprint{astro-ph/9911100}.

\bibitem[{\citenamefont{Kunze and Komatsu}(2014)}]{Kunze:2013uja}
\bibinfo{author}{\bibfnamefont{K.~E.} \bibnamefont{Kunze}} \bibnamefont{and}
  \bibinfo{author}{\bibfnamefont{E.}~\bibnamefont{Komatsu}},
  \bibinfo{journal}{JCAP} \textbf{\bibinfo{volume}{01}}, \bibinfo{pages}{009}
  (\bibinfo{year}{2014}), \eprint{1309.7994}.

\bibitem[{\citenamefont{Jedamzik and Saveliev}(2019)}]{Jedamzik:2018itu}
\bibinfo{author}{\bibfnamefont{K.}~\bibnamefont{Jedamzik}} \bibnamefont{and}
  \bibinfo{author}{\bibfnamefont{A.}~\bibnamefont{Saveliev}},
  \bibinfo{journal}{Phys. Rev. Lett.} \textbf{\bibinfo{volume}{123}},
  \bibinfo{pages}{021301} (\bibinfo{year}{2019}), \eprint{1804.06115}.

\bibitem[{\citenamefont{Lesgourgues}(2011)}]{Lesgourgues2011Cosmic}
\bibinfo{author}{\bibfnamefont{J.}~\bibnamefont{Lesgourgues}}
  (\bibinfo{year}{2011}), \eprint{1104.2932}.

\bibitem[{\citenamefont{Blas et~al.}(2011)\citenamefont{Blas, Lesgourgues, and
  Tram}}]{Blas2011Cosmic}
\bibinfo{author}{\bibfnamefont{D.}~\bibnamefont{Blas}},
  \bibinfo{author}{\bibfnamefont{J.}~\bibnamefont{Lesgourgues}},
  \bibnamefont{and} \bibinfo{author}{\bibfnamefont{T.}~\bibnamefont{Tram}},
  \bibinfo{journal}{JCAP} \textbf{\bibinfo{volume}{1107}}, \bibinfo{pages}{034}
  (\bibinfo{year}{2011}), \eprint{1104.2933}.

\bibitem[{\citenamefont{Audren et~al.}(2013)\citenamefont{Audren, Lesgourgues,
  Benabed, and Prunet}}]{Audren2013Conservative}
\bibinfo{author}{\bibfnamefont{B.}~\bibnamefont{Audren}},
  \bibinfo{author}{\bibfnamefont{J.}~\bibnamefont{Lesgourgues}},
  \bibinfo{author}{\bibfnamefont{K.}~\bibnamefont{Benabed}}, \bibnamefont{and}
  \bibinfo{author}{\bibfnamefont{S.}~\bibnamefont{Prunet}},
  \bibinfo{journal}{JCAP} \textbf{\bibinfo{volume}{1302}}, \bibinfo{pages}{001}
  (\bibinfo{year}{2013}), \eprint{1210.7183}.

\bibitem[{\citenamefont{Brinckmann and
  Lesgourgues}(2018)}]{Brinckmann2018MontePython}
\bibinfo{author}{\bibfnamefont{T.}~\bibnamefont{Brinckmann}} \bibnamefont{and}
  \bibinfo{author}{\bibfnamefont{J.}~\bibnamefont{Lesgourgues}}
  (\bibinfo{year}{2018}), \eprint{1804.07261}.

\bibitem[{\citenamefont{Beutler et~al.}(2011)\citenamefont{Beutler, Blake,
  Colless, Jones, Staveley-Smith, Campbell, Parker, Saunders, and
  Watson}}]{Beutler2011Galaxy}
\bibinfo{author}{\bibfnamefont{F.}~\bibnamefont{Beutler}},
  \bibinfo{author}{\bibfnamefont{C.}~\bibnamefont{Blake}},
  \bibinfo{author}{\bibfnamefont{M.}~\bibnamefont{Colless}},
  \bibinfo{author}{\bibfnamefont{D.~H.} \bibnamefont{Jones}},
  \bibinfo{author}{\bibfnamefont{L.}~\bibnamefont{Staveley-Smith}},
  \bibinfo{author}{\bibfnamefont{L.}~\bibnamefont{Campbell}},
  \bibinfo{author}{\bibfnamefont{Q.}~\bibnamefont{Parker}},
  \bibinfo{author}{\bibfnamefont{W.}~\bibnamefont{Saunders}}, \bibnamefont{and}
  \bibinfo{author}{\bibfnamefont{F.}~\bibnamefont{Watson}},
  \bibinfo{journal}{Mon. Not. Roy. Astron. Soc.}
  \textbf{\bibinfo{volume}{416}}, \bibinfo{pages}{3017} (\bibinfo{year}{2011}),
  \eprint{1106.3366}.

\bibitem[{\citenamefont{Ross et~al.}(2015)\citenamefont{Ross, Samushia,
  Howlett, Percival, Burden, and Manera}}]{Ross2014Clustering}
\bibinfo{author}{\bibfnamefont{A.~J.} \bibnamefont{Ross}},
  \bibinfo{author}{\bibfnamefont{L.}~\bibnamefont{Samushia}},
  \bibinfo{author}{\bibfnamefont{C.}~\bibnamefont{Howlett}},
  \bibinfo{author}{\bibfnamefont{W.~J.} \bibnamefont{Percival}},
  \bibinfo{author}{\bibfnamefont{A.}~\bibnamefont{Burden}}, \bibnamefont{and}
  \bibinfo{author}{\bibfnamefont{M.}~\bibnamefont{Manera}},
  \bibinfo{journal}{Mon. Not. Roy. Astron. Soc.}
  \textbf{\bibinfo{volume}{449}}, \bibinfo{pages}{835} (\bibinfo{year}{2015}),
  \eprint{1409.3242}.

\bibitem[{\citenamefont{Alam et~al.}(2017)}]{Alam2016Clustering}
\bibinfo{author}{\bibfnamefont{S.}~\bibnamefont{Alam}} \bibnamefont{et~al.}
  (\bibinfo{collaboration}{BOSS}), \bibinfo{journal}{Mon. Not. Roy. Astron.
  Soc.} \textbf{\bibinfo{volume}{470}}, \bibinfo{pages}{2617}
  (\bibinfo{year}{2017}), \eprint{1607.03155}.

\bibitem[{\citenamefont{Scolnic et~al.}(2018)}]{Scolnic2017Complete}
\bibinfo{author}{\bibfnamefont{D.~M.} \bibnamefont{Scolnic}}
  \bibnamefont{et~al.}, \bibinfo{journal}{Astrophys. J.}
  \textbf{\bibinfo{volume}{859}}, \bibinfo{pages}{101} (\bibinfo{year}{2018}),
  \eprint{1710.00845}.

\bibitem[{\citenamefont{André et~al.}(2014)}]{Andre2014Prism}
\bibinfo{author}{\bibfnamefont{P.}~\bibnamefont{André}} \bibnamefont{et~al.}
  (\bibinfo{collaboration}{PRISM}), \bibinfo{journal}{JCAP}
  \textbf{\bibinfo{volume}{1402}}, \bibinfo{pages}{006} (\bibinfo{year}{2014}),
  \eprint{1310.1554}.

\bibitem[{\citenamefont{Perotto et~al.}(2006)\citenamefont{Perotto,
  Lesgourgues, Hannestad, Tu, and Wong}}]{Perotto2006Probing}
\bibinfo{author}{\bibfnamefont{L.}~\bibnamefont{Perotto}},
  \bibinfo{author}{\bibfnamefont{J.}~\bibnamefont{Lesgourgues}},
  \bibinfo{author}{\bibfnamefont{S.}~\bibnamefont{Hannestad}},
  \bibinfo{author}{\bibfnamefont{H.}~\bibnamefont{Tu}}, \bibnamefont{and}
  \bibinfo{author}{\bibfnamefont{Y.~Y.~Y.} \bibnamefont{Wong}},
  \bibinfo{journal}{JCAP} \textbf{\bibinfo{volume}{0610}}, \bibinfo{pages}{013}
  (\bibinfo{year}{2006}), \eprint{astro-ph/0606227}.

\bibitem[{\citenamefont{Brinckmann et~al.}(2019)\citenamefont{Brinckmann,
  Hooper, Archidiacono, Lesgourgues, and Sprenger}}]{Brinckmann2018Promising}
\bibinfo{author}{\bibfnamefont{T.}~\bibnamefont{Brinckmann}},
  \bibinfo{author}{\bibfnamefont{D.~C.} \bibnamefont{Hooper}},
  \bibinfo{author}{\bibfnamefont{M.}~\bibnamefont{Archidiacono}},
  \bibinfo{author}{\bibfnamefont{J.}~\bibnamefont{Lesgourgues}},
  \bibnamefont{and} \bibinfo{author}{\bibfnamefont{T.}~\bibnamefont{Sprenger}},
  \bibinfo{journal}{JCAP} \textbf{\bibinfo{volume}{1901}}, \bibinfo{pages}{059}
  (\bibinfo{year}{2019}), \eprint{1808.05955}.

\bibitem[{\citenamefont{Kogut et~al.}(2011)}]{Kogut2011Primordial}
\bibinfo{author}{\bibfnamefont{A.}~\bibnamefont{Kogut}} \bibnamefont{et~al.},
  \bibinfo{journal}{Journal of Cosmology and Astro-Particle Physics}
  \textbf{\bibinfo{volume}{2011}}, \bibinfo{eid}{025} (\bibinfo{year}{2011}),
  \eprint{1105.2044}.

\bibitem[{\citenamefont{Abitbol et~al.}(2017)\citenamefont{Abitbol, Chluba,
  Hill, and Johnson}}]{Abitbol2017Prospects}
\bibinfo{author}{\bibfnamefont{M.}~\bibnamefont{Abitbol}},
  \bibinfo{author}{\bibfnamefont{J.}~\bibnamefont{Chluba}},
  \bibinfo{author}{\bibfnamefont{C.}~\bibnamefont{Hill}}, \bibnamefont{and}
  \bibinfo{author}{\bibfnamefont{B.}~\bibnamefont{Johnson}},
  \bibinfo{journal}{Mon. Not. Roy. Astron. Soc.}
  \textbf{\bibinfo{volume}{471}}, \bibinfo{pages}{1126} (\bibinfo{year}{2017}),
  \eprint{1705.01534}.

\bibitem[{\citenamefont{Douspis et~al.}(2018)\citenamefont{Douspis, Salvati,
  and Aghanim}}]{Douspis:2018xlj}
\bibinfo{author}{\bibfnamefont{M.}~\bibnamefont{Douspis}},
  \bibinfo{author}{\bibfnamefont{L.}~\bibnamefont{Salvati}}, \bibnamefont{and}
  \bibinfo{author}{\bibfnamefont{N.}~\bibnamefont{Aghanim}},
  \bibinfo{journal}{PoS} \textbf{\bibinfo{volume}{EDSU2018}},
  \bibinfo{pages}{037} (\bibinfo{year}{2018}), \eprint{1901.05289}.

\bibitem[{\citenamefont{Bolliet}(2018)}]{Bolliet:2018yaf}
\bibinfo{author}{\bibfnamefont{B.}~\bibnamefont{Bolliet}}, in
  \emph{\bibinfo{booktitle}{{53rd Rencontres de Moriond on Cosmology}}}
  (\bibinfo{year}{2018}), pp. \bibinfo{pages}{165--169}, \eprint{1806.04786}.

\end{thebibliography}

\end{document}